\documentclass[twocolumn, amsmath, superscriptaddress, amsfonts,prb]{revtex4}

\usepackage{graphicx}
\usepackage{lipsum}

\begin{document}

\title{Gamma- and x-ray accelerated tin whisker development}
    \author{Osama Oudat}\affiliation{Department of Physics and Astronomy, University of Toledo, Toledo,OH 43606, USA}
\author{Vidheesha Arora}\affiliation{Department of Physics and Astronomy, University of Toledo, Toledo,OH 43606, USA}
\author{E. Ishmael Parsai} \affiliation{Department of Radiation Oncology, University of Toledo Health Science Campus, Toledo, OH 43614, USA}
\author{Victor G. Karpov}\affiliation{Department of Physics and Astronomy, University of Toledo, Toledo,OH 43606, USA}
\author{Diana Shvydka}\email{diana.shvydka@utoledo.edu}\affiliation{Department of Radiation Oncology, University of Toledo Health Science Campus, Toledo, OH 43614, USA}
\begin{abstract}
 We observed accelerated whisker development in thin tin films under non-destructive gamma-ray and x-ray irradiation sources. The effect is mediated by charges induced in glass substrates supporting films, and becomes significant reaching the characteristic range of radiation doses of 20-30 kGy. We were able to change the radiation induced whisker growth rate by electrically disconnecting some parts of our experimental setup thus demonstrating the electrostatic nature of the whisker development. The observed acceleration factors make the ionizing radiation a potential non-destructive and readily implementable accelerated life testing tool.
\end{abstract}

\maketitle

\section{Introduction}\label{sec:intro}

Spontaneous growth of thin hair-like protrusions with high aspect-ratios, called whiskers, was observed for many technologically important metals, such as tin, zinc, cadmium, and others. \cite{NASA,brusse} Whiskers present a serious reliability problem due to their caused random electrical shorts in a wide range of devices from consumer microelectronics to military and space applications. Despite significant research effort devoted to the problem to this day there is no commonly accepted mechanism of whisker development. That lack of understanding makes it difficult to develop reliable accelerated life testing procedures that, while direly needed in industry, remain disputable.\cite{JEDEC201,NASAcom}

A recently proposed electrostatic theory \cite{karpov2014,karpov2015} offers a unique mechanism of whisker development with predictions encompassing many whisker observations. Moreover, based on that theory of electrostatic interactions, one can design certain approaches to  accelerated testing in strong electric fields. Such field-induced accelerated whisker growth has been observed in multiple experiments. \cite{vasko2015a,borra2016,niraula2016}

A related effort was devoted to acceleration of whisker growth under ionizing radiation sources, such as 6 MeV electrons \cite{niraula2016,vasko2015} and 0.4 MeV gamma-rays \cite{Killefer} capable of inducing electric fields by electrically charging a substrate (glass or acrilic holder) supporting tin or zinc films. We note that radiation effects on whisker growth have been observed $\sim$60 years ago and remained unexplained.\cite{ellis1958}

Here, we further explore the effects of ionizing radiation on tin whisker development with a goal of verifying the hypothesis of their underlying electrostatic interaction. A central argument will be that while the utilized radiation fields cannot lead to metal film structure modifications, they accelerate tin whisker development by orders of magnitude; hence, ionization and its related substrate charging \cite{shaneyfelt,schwank,vasko2015,Killefer} becomes the culprit.  In order to verify that hypothesis, we studied the dependence of whisker development rates on the ionization rate by designing experiments where samples received different radiation intensities. Furthermore, to verify the role of electrostatic interactions we implemented a design where a part of the sample was screened against radiation yet remaining electrically connected to the exposed part. To describe the electrostatic effects for the case of laterally nonuniform charge generation (including partial sample screening) we solved its related electrostatic problem and compared the results to our experimental data.

Our paper is organized as follows. In Sec. \ref{sec:methods} we provide the details of the experiments and statistical analysis carried out to generate the results. In that same section, we present a mathematical solution to the electrostatic problem of field generation in the electrically connected system with laterally non-uniform charge generation. Sec. \ref{sec:results}  details the main findings and outcomes of our study. The significance of our results and comparison with previous work is presented in Sec. \ref{sec:disc}. Finally, Sec. \ref{sec:concl} lists the main conclusions of our study, highlighting the electrostatic nature of observed effects and their practical implications.

\section{Methods}\label{sec:methods}

\subsection{Samples}\label{sec:samples}

\begin{figure}[hbt]
\centering
\includegraphics[width=0.37\textwidth]{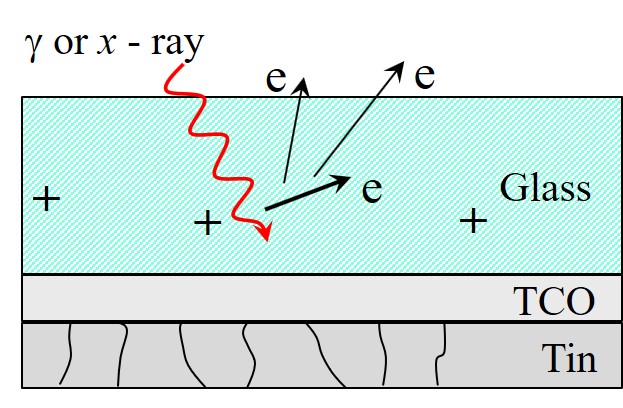}
\caption{A sketch of the sample structure, glass/TCO/polycrystalline Sn (not to scale), under irradiation entering from the substrate (glass) side. The substrate becomes charged due to generated secondary electrons. \label{fig:TCO_Sn}}.
\end{figure}

Thin-film Sn samples were deposited on 3-mm thick soda-lime glass substrates coated with transparent conducting oxide (TCO, specifically, SnO$_2$:F with nominal 15 Ohm/square sheet resistance; TEC-15 glass from Pilkington), illustrated in Fig. \ref{fig:TCO_Sn}. This type of substrate supports good film adhesion with various deposition applications. It was proven earlier that Sn films deposited on such substrates can grow whiskers at hight enough rates. \cite{vasko2015,vasko2015a,borra2016,niraula2016} We followed our standard protocol where substrates cut into 3x6 cm$^2$ pieces are washed in cleaning solution (Micro-90), then rinsed with de-ionized water (DI), followed with ultra-sonication bath in acetone for 25 min, and, finally, with ultra-sonication bath in DI for 25 min. In between these steps, the surfaces are blown dry with nitrogen.

The following experiment was designed to verify the electric nature of whisker driving force. Fresh Sn samples were prepared by vacuum evaporation of 99.999\% pure Sn (from Kurt J. Lesker) from tungsten boats; the detailed fabrication process steps are described in Ref. \onlinecite{borra2016}. Sn layers of 250$\pm$25nm thickness were deposited at a growth rate of 2-4 \AA/s through stainless steel masks, resulting in a set of 5 mm wide Sn strips  separated by of $\sim 1$cm distance, as illustrated in Fig. \ref{fig:Strips}, over continuous TCO coating. Each Sn strip was electrically disconnected with a set of parallel isolation TCO-eliminating scribes; this was confirmed through measurement of infinite resistance between the isolated parts.

For comparison, some tin films were deposited on a bare glass surface opposite to TCO to be exposed to the same type of radiation.

We also used a set of `mature' Sn thin-film samples, about the same thickness, rf-sputtered on the same TEC-15 glass substrates 3 years prior, \cite{vasko2015} already showing significant whisker growth before irradiation. For both fresh and "mature" samples the Sn film thickness was far below the typical micron range of efficient whisker growing
film thickness. \cite{NASA}

\subsection{Radiation sources and design}\label{sec:radiation}

\begin{figure}[hbt]
\centering
\includegraphics[width=0.4\textwidth]{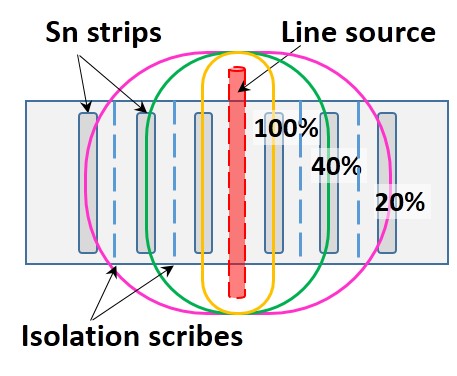}
\caption{A sketch of the gamma-ray irradiation geometry: line source located in the middle of the sample (red dashed cylinder); a dose fall-off from a line source follows 1/r law, shown with labeled isodose lines at locations of metallized Sn strips. All strips were electrically disconnected along the shown isolation scribes (additional central scribe coincides with the position of the source). \label{fig:Strips}}.
\end{figure}

Two types of radiation sources were utilized: lower energy gamma-rays and higher energy x-rays. We note that both gamma- and x-rays are sources of photons, the distinction is made based on their origin correspondingly from either nuclear disintegrations (radioactive decay) or high-energy electron interactions with nuclei (bremsstrahlung).
For irradiation of a fresh Sn film sample we used gamma-ray source of Ir-192 clinical high-dose rate (HDR) afterloader (VariSource, Varian Medical Systems). That source is poly-energetic, with the average photon energy $\sim$380keV (beta particles emitted by the source are fully absorbed by Ni-Ti alloy encapsulating the radionuclide core), and $\sim$74 days half-life. \cite{borg} The corresponding photon attenuation in a very thin Sn film is negligible (interaction probability $\sim10^{-7}$), while the glass substrate attenuates a few percent of the photon beam over its 0.3 cm thickness. \cite{NIST}
The source of linear dimension of $\sim$3 mm, was used at multiple dwell positions 5 mm apart along a straight line, which makes it effectively a line source positioned through the center of the sample, parallel to its shorter dimension, as shown in Fig. \ref{fig:Strips}.

Samples were irradiated to dose levels of 10, 20 and 30 kGy (kGy=$10^3$ Gy, where 1 Gy=1 J/kg, "gray", is SI unit of radiation dose) at the central strips location. The distances from the strips to the gamma-ray source (of linear dimension of $\sim$3 mm) were significantly different ranging from $\sim 1$ to $\sim 3$ cm, so the strips received the correspondingly different doses, following $1/r$ dose fall-off with the distance $r$, as illustrated in Fig. \ref{fig:Strips}. The required HDR source dwell times were calculated using BrachyVision treatment planning system. The control sample was stored under controlled lab conditions for the experiment duration, with the exception of being periodically transported for imaging.

The dose was delivered in multiple irradiation sessions of 2-4 hours per 1 kGy. It was verified in our previous work \cite{vasko2015} that sample electrization decayed very fast upon cessation of irradiation. Hence, the integral time of electric field exposure was calculated as the sum of that of irradiation sessions time intervals (see also the discussion in Sec. \ref{sec:disc}).


\begin{figure}[hbt]
\centering
\includegraphics[width=0.45\textwidth]{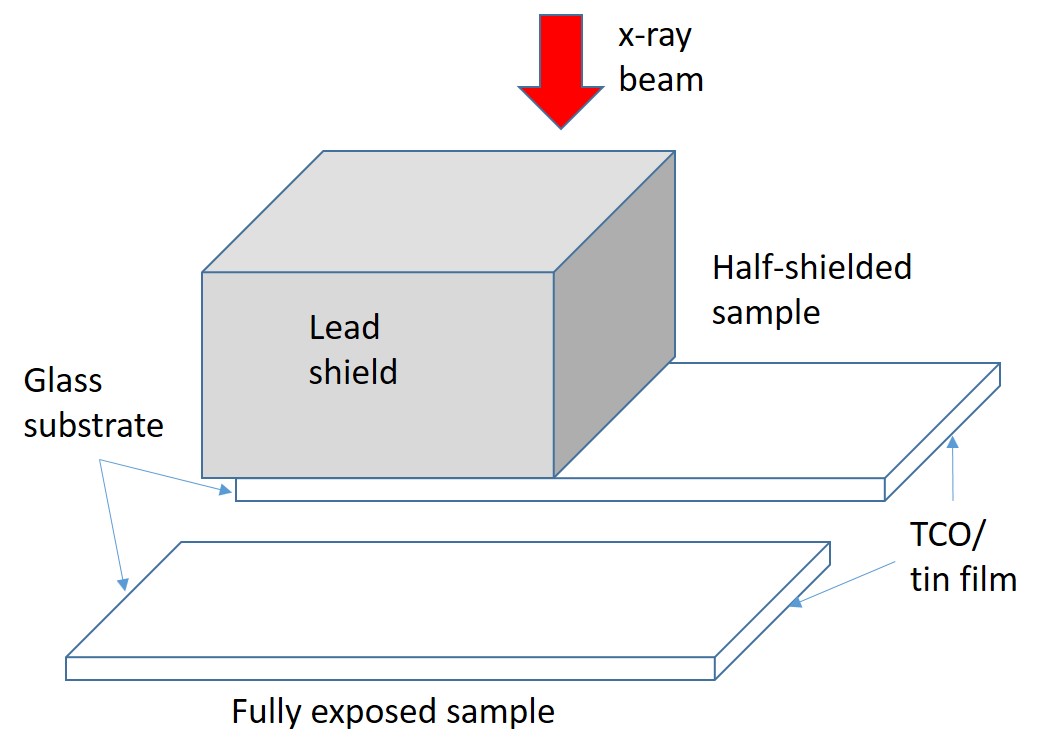}
\caption{A sketch of the x-ray irradiation geometry with shielding: the photons are coming from the top down. They can be shielded by a thick lead block for a half of a sample, while another sample is left unshielded. \label{fig:shield}}.
\end{figure}

To further elucidate the role of photon-induced electric field we conducted an experiment in which half of the irradiated sample was shielded with lead to bring the photon intensity down to less than 1/10th of the unobstructed source. Since use of the gamma-ray source was not feasible for such an experiment due to prohibitively long irradiation times, we used a significantly higher dose rate source of a medical linear accelerator (linac) producing x-ray spectrum with the maximum energy of 10 MeV and the average energy close to 1.5 MeV (a flattening-filter free, or 10FFF, source).

A 10FFF photon beam of a Varian Edge linac with the dose rate of 24 Gy/min at the reference depth was used for a partial shielding experiment. Samples were placed glass substrate side up at the closest possible position from the source ($\sim $ few cm from the gantry to accommodate the lead blocks used for partial shielding) to further increase the dose rate; the irradiation geometry is illustrated in Fig. \ref{fig:shield}. This resulted in irradiation times of 3.2 hours to achieve 10 kGy dose in glass substrate of the fully irradiated sample (compare to 20 to 40 hours required with the gamma-ray source of the HDR unit). Two samples were placed under the beam: one fully exposed and one half-shielded, where the lead shield of the total thickness of 8 cm was used to cover half of the sample (a tenth-value layer of lead under a typical 10 MV linac beam \cite{Khan} is $\sim 6$ cm) while exposing the other half to the same beam intensity as the fully irradiated sample. The control sample was stored under lab conditions for the duration of experiments.

For this round of experiments we used the `mature' Sn thin-film samples with significant whisker coverage before irradiation (0kGy). Whisker statistics for this experiment was collected similarly to the set irradiated under gamma-rays. All three samples, control, fully irradiated, and half-shielded (shielded and not shielded halves analyzed separately) were imaged before irradiations, or at 0 kGy, and after 10, 20, and 30 kGy dose levels were achieved for irradiated samples.

Here we address a question of the probability of atomic displacements in Sn film, and related potentially damaging effect of our high-energy radiation sources, in the unlikely event of photon absorption in the film itself. The primary photon interaction with samples for both sources is Compton scattering, resulting in the energy transfer to an electron, set in motion after the interaction. For the utilized  gamma-ray spectrum (the average energy of 0.38 MeV and maximum energy below 1 MeV)\cite {borg}. The average energy received by Compton electron is approximately 40\% of the source photon energy, \cite{attix} and is never higher than its maximum energy. The energy transferred to an atom is smaller than the electron energy by the ratio of the electron to atom masses, $\sim 4\times 10^{-6}$. As a result, the atom receives $<$1eV on average and cannot receive more than 4eV, a value  well below the displacement threshold energy estimated for Sn as 22$\pm 2$ eV. \cite{mcilwan1975,andersen1979} Other mechanisms of energy transfer are even less efficient. \cite{oen} For the x-ray source (the average energy of 1.5 MeV and maximum energy 10 MeV) the average electron energy received is approximately 50\% of the source photon energy, leading again to $<$1eV on average, but this time $\sim 40$ eV maximum energy transferred to an atom. While the latter estimate makes atomic displacements possible in principle, their probability is  vanishingly small. Thereby the radiation sources used do not produce detectable atomic displacements, and are suitable for non-destructive whisker propensity testing.

\subsection{Imaging}\label{sec:imaging}

\begin{figure}[bh]
\centering
\includegraphics[width=0.45\textwidth]{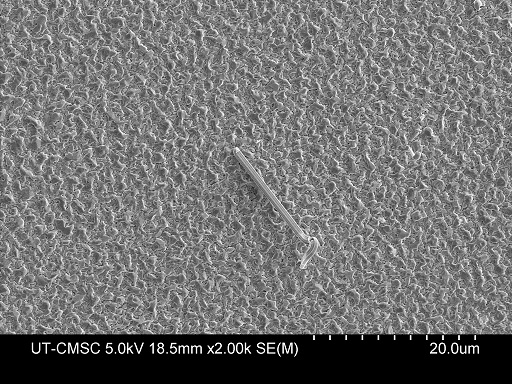}
\caption{An SEM image of the longest observed whisker for the fresh Sn sample irradiated to 30 kGy under x-ray source, central strip. \label{fig:images}}.
\end{figure}

A scanning electron microscope (SEM), Hitachi S-4800, was utilized as a primary characterization tool for metal film surface imaging. It was operated in secondary-electron mode with acceleration voltage of 5 kV in order to limit the observations to the film surface. SEM imaging of the irradiated samples were conducted by the following schedule: before irradiation (0 kGy), and after receiving 10, 20, and 30kGy total dose. The control sample was imaged together with irradiated samples receiving their next 10kGy dose, on average once a week; the overall experiment duration was $\sim30$ days.

For each sample we collected 40 SEM images per dose level (0, 10, 20, and 30kGy), imaging areas were randomly selected for each imaging session. Whiskers were counted in all collected images, and the ImageJ software package was used to measure the lengths of all metal whiskers identified on sample surfaces. The whisker length was measured as a sum of straight portions for whiskers with complex shapes.

For a gamma-ray irradiated sample we processed images from three different regions, based on their proximity to the source and the average dose level received (100\%, 40\%, or 20\% of the total dose, Fig. \ref{fig:Strips}). A representative SEM image of the area with the longest observed whisker is shown in Figure \ref{fig:images} taken after 30kGy dose delivered to a strip from the central region.

\subsection{Theoretical methods}\label{sec:theory}

\begin{figure}[h]
\includegraphics[width=0.40\textwidth]{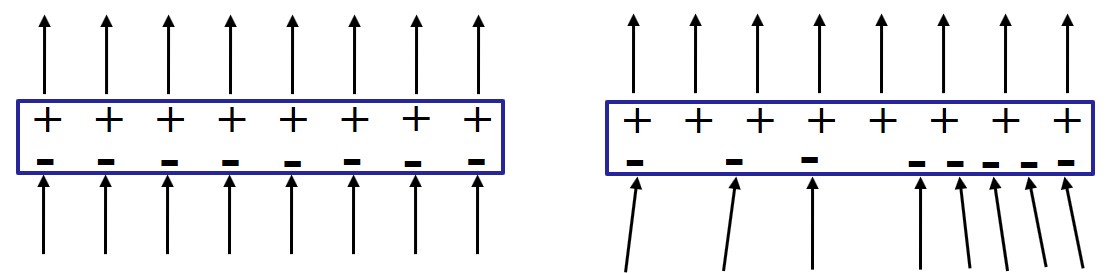}
\caption{A sketch of the TCO charge carrier distribution in response to the electric field of the underlying substrate. The free carriers are chosen to be negative; the positive background remains intact. Left: uniform substrate field. Right: a nonuniform field causing the free carrier redistribution. \label{fig:screening}}.
\end{figure}

When thin TCO layer is put on top of a substrate containing a nonuniform charge density, its free charge carriers can non-uniformly rearrange themselves thus partially screening the substrate generated field as illustrated in Fig. \ref{fig:screening}. Here we consider that effect more quantitatively to show how that screening cancels the nonuniform part not affecting the average (uniform) component of the field. For simplicity, we assume a one-dimensional lateral nonuniformity in substrate charge density, $g(y)$.

\begin{figure}[t!]
\includegraphics[width=0.45\textwidth]{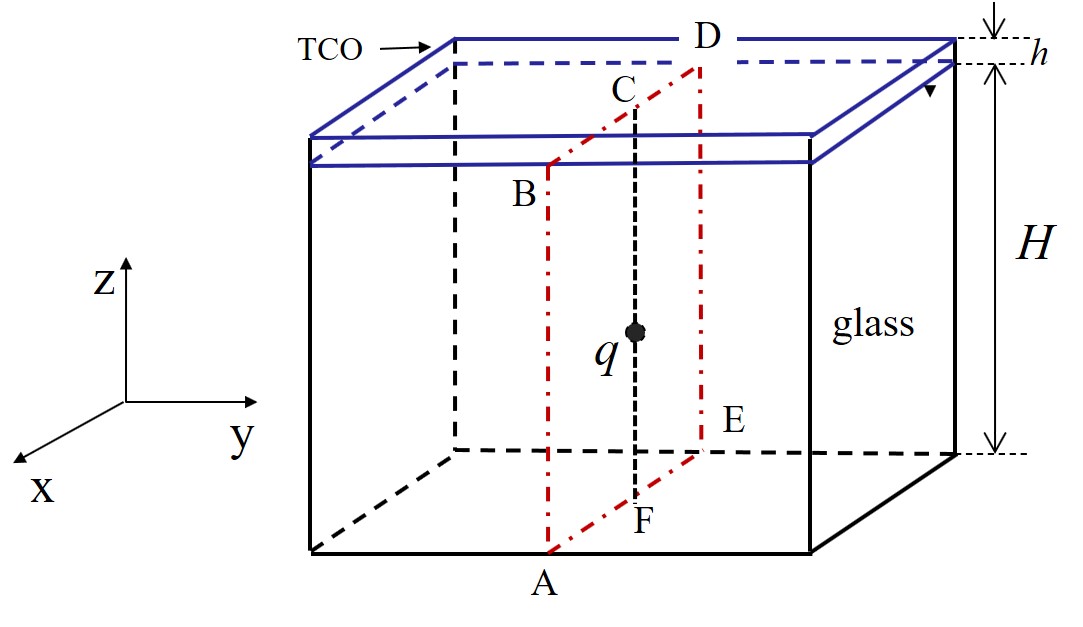}
\caption{A geometry of the substrate-TCO system showing a point charge $q$ generating the field in TCO, and the charge plane ABCD composed of such charges.   \label{fig:geometry}}.
\end{figure}

We present the TCO charge density as $\rho _{av}+\rho (y)$ thus separating the average values and the nonuniform components, whose integral value is zero, $\int\rho (y)dy=0$.
The shape $\rho (y)$ is determined by the condition that the tangential component of the electric field is zero. We explore the latter condition starting with the case of a point charge  $q$ in the substrate (see Fig. \ref{fig:geometry}). By replacing $q$ with an elemental charge $\rho (y)dxdydz$ and integrating in $z,y$ plane yields the surface
charge density induced by the plate ABCD of thickness $dy$ at a given $y$. We take into account that the TCO thickness $h$ is much smaller than that of substrate, $h\ll H$.


We use the standard result for the surface charge density induced by a point charge in a metal surface,
\begin{equation}\label{eq:point}
\rho _q(x,y,z)=-\frac{zq}{2\pi (x^2+y^2+z^2)^{3/2}}.\end{equation}
Setting $q\rightarrow g(y)dxdydz$ where $g$ is the charge density in the substrate, and integrating over $z$ from 0 to $H$ and over $x$ from $-\infty$ to $\infty$
yields the surface charge density induced by the flat plate ABCD of thickness $dy$ at distance $y$,
\begin{equation}\label{eq:ABDE}\rho _{ABCD}=-\frac{1}{2\pi}\ln\left(\frac{y^2+H^2}{y^2}\right)gdy
\end{equation}

From Eq. (\ref{eq:ABDE}), the TCO charge density distance $y$ from the center of the sample of length $L$ (in $y-$direction) is given by,
\begin{equation}\label{eq:final}
\rho (y)=-\frac{1}{2\pi}\int _{-L/2}^{L/2}\ln\left[\frac{(y'-y)^2+H^2}{(y'-y)^2}\right]g(y')dy'.
\end{equation}
where $L$ is the length (along $y$-axis) of the sample.
According to the Appendix, the integrand is reduced to the delta function, $2\pi H\delta (y'-y)$ when $H\ll L$, yielding

\begin{equation}\label{eq:rho}
\rho (y)=-g(y)H
\end{equation}
Therefore, the nonuniform part of the screening surface charge density $\rho (y)$ is the same in absolute value and opposite in sign to the substrate non-uniformity canceling exactly the substrate nonuniform charge density $g(y)H$. We conclude that the nonuniform component of the electric field is shielded completely by the TCO layer.

The latter derivation and conclusion assumes electrically neutral conductive TCO layer, so that $\int \rho (y)dy =-\int g(y)dy=0$. An alternative condition can be that of the grounded TCO allowing charge accumulation and total screening of the substrate charge. That latter condition is of no interest here as not corresponding to the experimental setup used in this study.

\section{Results}\label{sec:results}

As a general observation, exposure to radiation increased whisker propensity in all of our experiments. We should note however that the radiation stimulated whisker development revealed itself upon a certain delay of weeks since the radiation exposure. A similar phenomenon was noticed in the pioneering work \cite{ellis1958} and our previous research. \cite{Killefer}

Using setup of Fig. \ref{fig:Strips} without electrically insulating scribes between the strips (continuous TCO layer), we observed whisker density and length not correlating with the proximity to the radiation source, but dependent only on the total irradiation dose received by the sample, similar to our previous experiment. \cite{Killefer} SEM scans of samples before irradiation, conducted within a week from sample deposition, revealed no whiskers; whiskering was first observed after irradiation to 20kGy, further enhanced after the final dose of 30 kGy was delivered.

It would be natural to assume that the non-uniform irradiation of the Fig. \ref{fig:Strips} setup should result in the correspondingly nonuniform distribution of whiskers, seemingly contradicting their observed uniform generation. That contradiction is resolved by assuming the electric field dominated whisker development and taking into account the field screening due to the electrons in the TCO layer below Sn strip regions. The electrostatic calculation in Sec. \ref{sec:theory} shows indeed that free charge carriers in TCO rearrange themselves to fully screen the nonuniform field component while not affecting the average (uniform) field perpendicular to the substrate.  We relate the latter to the observed significant acceleration of whisker growth.

\begin{figure}[b!]
\centering
\includegraphics[width=0.5\textwidth]{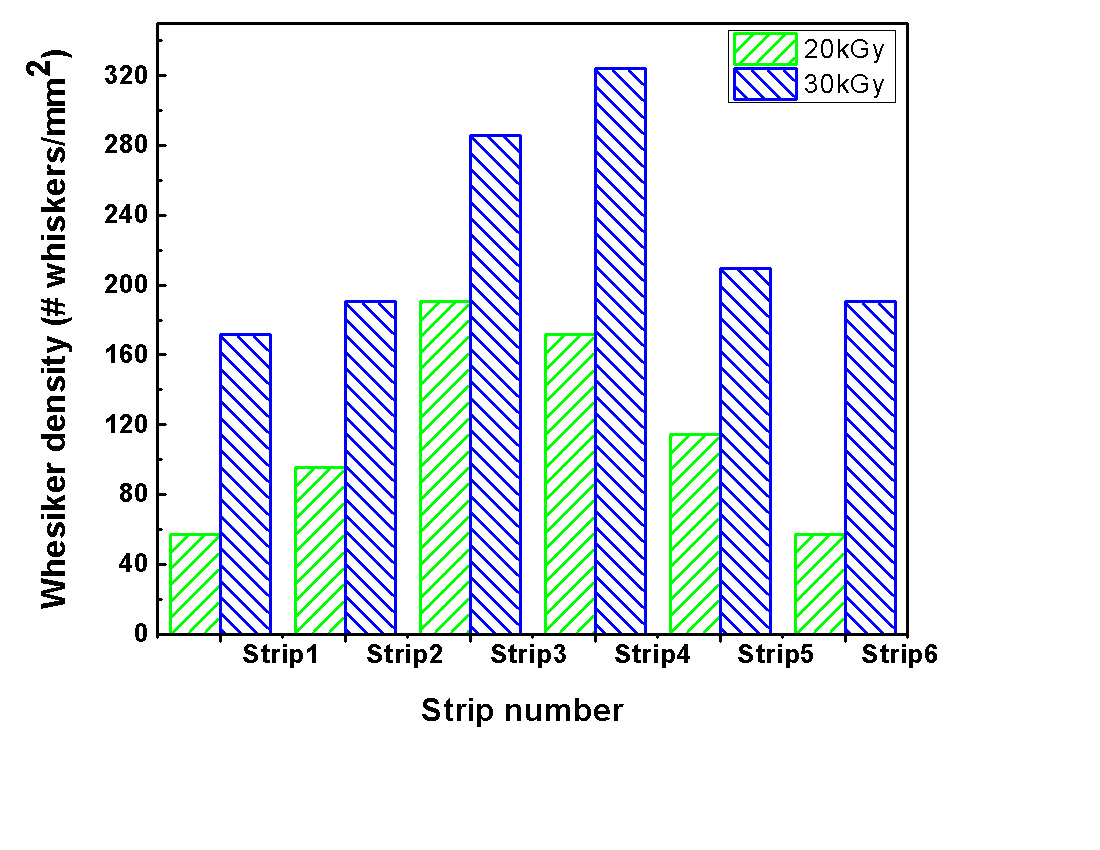}
\caption{Summary whisker statistics for the sample irradiated under gamma-rays to 20 and 30kGy doses, grouped by location on a corresponding electrically insulated Sn film strip (0kGy here corresponds to no whiskers in a freshly deposited sample). Frequency count for whisker density, 1/mm$^2$.\label{fig:statsHDR}}.
\end{figure}

\begin{figure}[h!]
\centering
\includegraphics[width=0.5\textwidth]{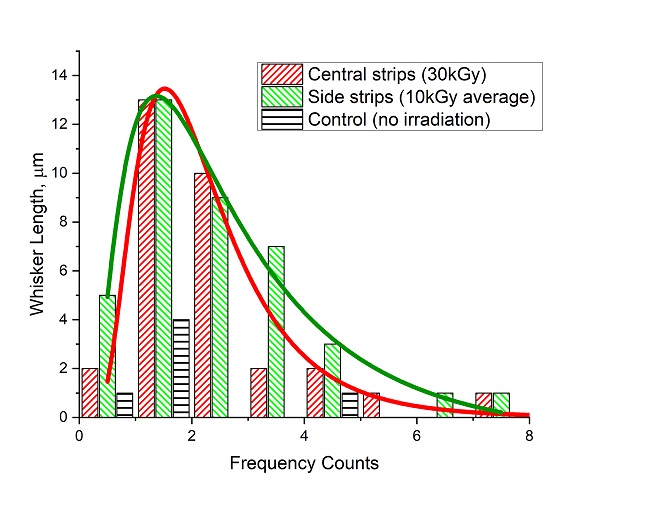}
\caption{Whisker length statistics: central region irradiated to 30kGy, side regions irradiated to the average dose of 10kGy. Data for control sample evaluated at the same time is included. The curves correspond to the log-normal fits.\label{fig:WL_HDR}}.
\end{figure}

\begin{figure}[b!]
\includegraphics[width=0.50\textwidth]{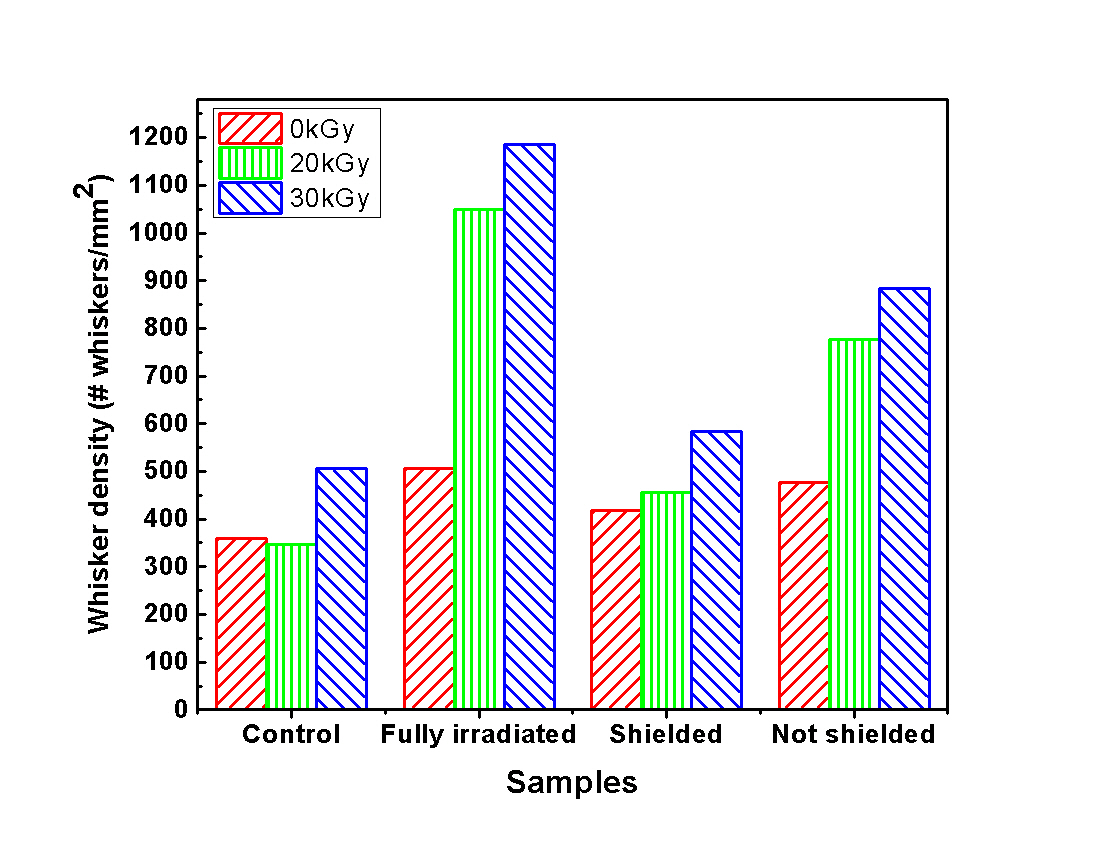}
\caption{Summary of the whisker density statistics for samples irradiated under 10FFF x-ray beam. Control sample, Fully irradiated, and Half-shielded (represented by two data sets labeled as `Shielded' and `Not shielded'). Image analysis at 0, 20, and 30kGy of radiation dose, corresponding to initial, and weeks 3 and 4 time intervals for the control sample.  \label{fig:stats10FFF}}.
\end{figure}

A whisker statistics summary for gamma-ray irradiations obtained with the setup of Fig. \ref{fig:Strips} {\it with electrically insulating scribes} between the strips is presented in Fig. \ref{fig:statsHDR} in a form of `frequency counts'. We observed whisker growth significantly accelerated under irradiation, with both whisker average density and length affected. The latter is illustrated in Fig. \ref{fig:WL_HDR} accompanied by the log-normal fits standard of tin whisker statistics. \cite{niraula2015} Side regions (2 strips furthest from the source on each side, 4 strips total) received about one-third of the dose delivered to the central region (1 strip closest to the source on each side, 2 strips total). We grouped sample statistics based on those two locations, which demonstrated noticeably different growth rates.  (A sample of the longest observed whisker is shown in Fig. \ref{fig:images} against the background of the thin film polycrystalline texture.) While the control sample also grew some whiskers, producing density of 66.7 whiskers/mm$^2$ by the final imaging session (when 30kGy dose level was reached at the central strips location), most of them were order of magnitude shorter, almost at the level of `nodules'. These observations are fully consistent with the hypothesis of electric field dominated whisker development where the field strength correlates with the radiation intensity received.

For the x-ray `shielding' experiment of Fig. \ref{fig:shield}, our results are presented in Fig. \ref{fig:stats10FFF}. Again, four radiation dose levels were incrementally applied:  0 kGy (i.e. before exposure), 10 kGy, 20 kGy, and 30 kGy (we omitted 10kGy dose results as not leading to statistically significant changes compared to 0 kGy). Each increment took certain time determined mostly by the equipment availability. Therefore, by referring to those 4 doses, we simultaneously refer to the 4 time instances when whisker imaging was performed.

The above 3 doses/times apply to 4 nominally identical samples used. They are defined as follows. The control sample presented in Fig.\ref{fig:stats10FFF} by the group of  measurements was never exposed to radiation. However its whisker concentration was measured 4 times upon the completion of the initial, 0 kGy scan, and 3 additional stages of irradiation of other samples.

The `Fully irradiated' sample went through all radiation exposures without any shielding and 4 examinations of its whisker concentration were performed. Finally, as depicted in Fig. \ref{fig:shield}, yet another sample was partially shielded while placed under the same radiation beam. Its irradiated and shielded halves were counted as a separate sample corresponding to two groups of measurements each shown in Fig. \ref{fig:stats10FFF}.

We recall that all the samples in this experiment were not freshly fabricated (`mature') and had a certain concentration of whiskers grown prior to the experiment. Our interpretation of the data in Fig. \ref{fig:stats10FFF} is presented in Sec. \ref{sec:disc} next.

\begin{figure}[hbt]
\centering
\includegraphics[width=0.35\textwidth]{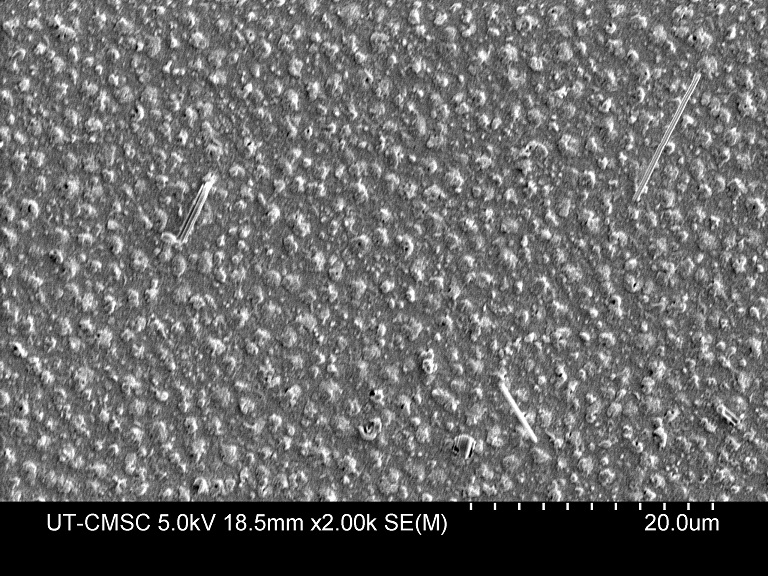}
\includegraphics[width=0.35\textwidth]{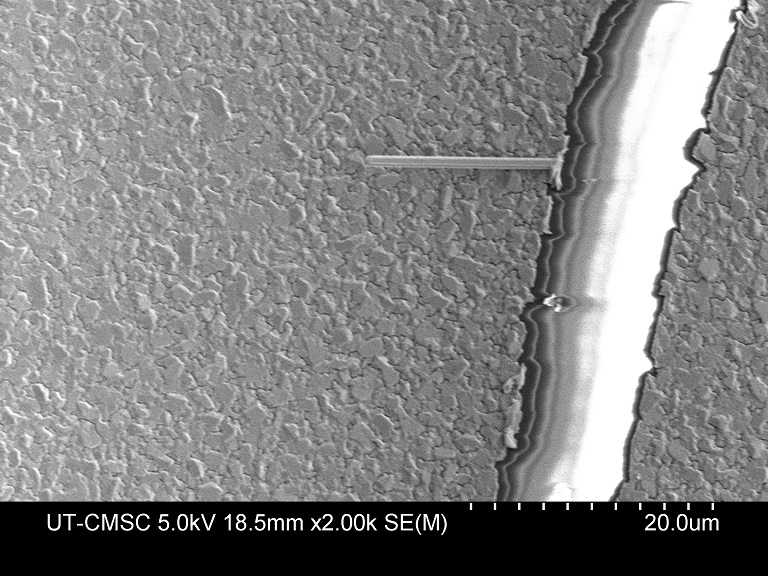}
\caption{Representative SEM images of Sn film deposited on bare glass with whiskers growing under gamma-ray irradiation. Bright strip in the bottom image shows the bare glass region where the film pilled-off with a whisker growing at the film edge. \label{fig:Glass_Sn}}.
\end{figure}

Finally, as a `side' result, we would like to mention our observation of whiskers on the irradiated tin films deposited on the bare glass surfaces mentioned in Sec. \ref{sec:methods} and illustrated in Fig. \ref{fig:Glass_Sn}. While we did not systematically studied such samples here, that observation may be of interest as demonstrating the versatility of substrate types compatible with tin whisker growth.

\section{Discussion}\label{sec:disc}

The data in Fig. \ref{fig:statsHDR}  indicate that electrically disconnected tin strips grow whiskers in concentrations correlated with the radiation dose received and varying between the strips as opposed to the earlier studied case of the strips electrically connected by continuous conducting TCO layer without scribes. \cite{Killefer} That fact is consistent with our theoretical methodology of Sec. \ref{sec:theory} showing how a laterally nonuniform substrate charging results in a laterally uniform electric field due to the free carriers rearrangement in TCO. The whisker length statistics in Fig. \ref{fig:WL_HDR} shows typically observed distribution  well-fitted with log-normal functions. Along with the fact that the radiation used was too soft for structural transformations and could only ionize the material, these data testify in favor of the electrostatic dominated whisker growth.

The data of Fig. \ref{fig:stats10FFF} can be attempted along similar lines. The half shielded sample will acquire a half of substrate charges compared to the fully irradiated one. Following the results of Sec. \ref{sec:theory}, its generated field will be determined by the {\it average} charge concentration, which because of the half shielding, will be two times lower than that of the fully exposed sample. We therefore expect that the field generated in both the unshielded and shielded halves (belonging to the same sample) should be two times weaker than that of the fully exposed sample.

\begin{figure}[htb]
\includegraphics[width=0.50\textwidth]{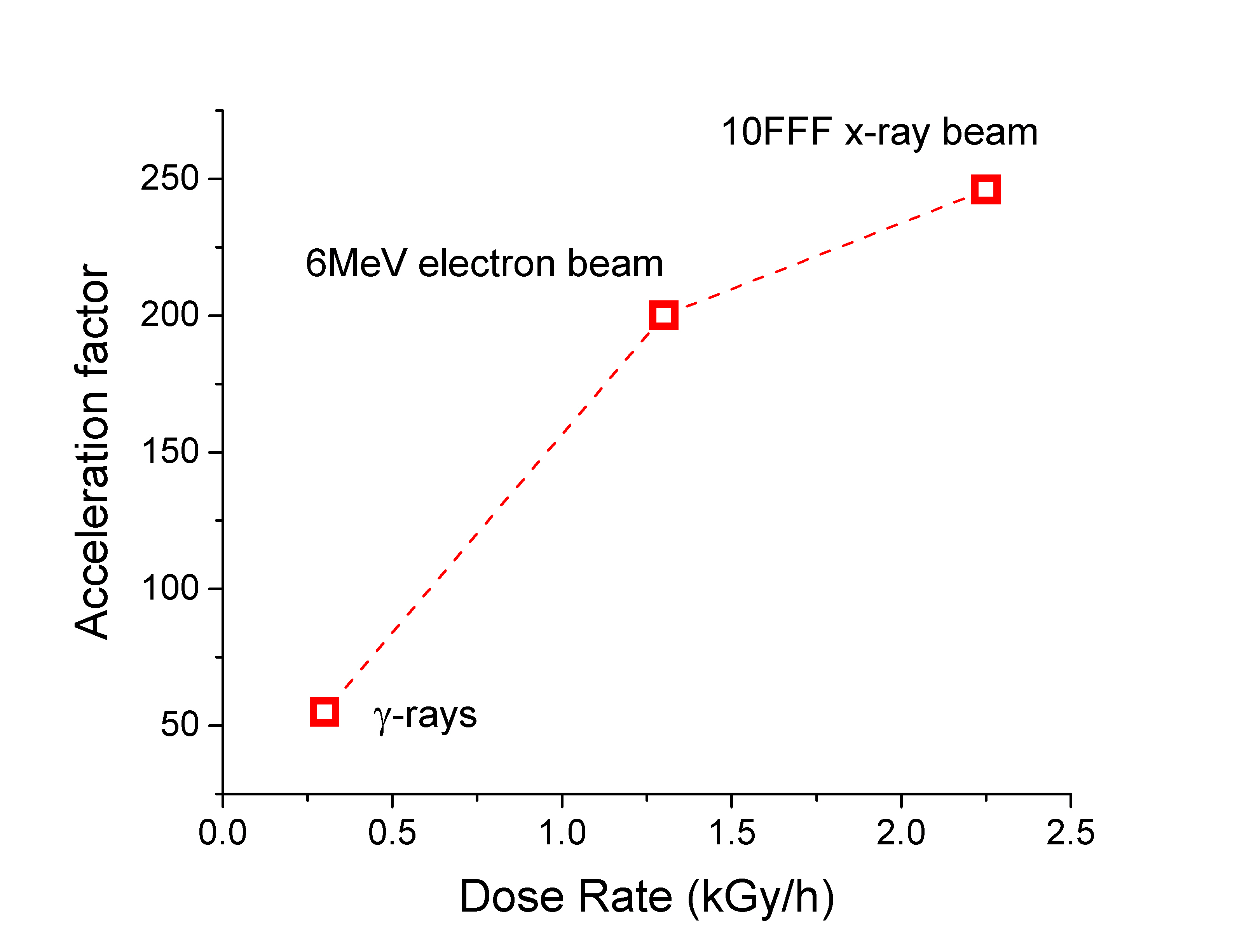}
\caption{Dose rate dependence of the acceleration factors calculated based on whisker density statistics for Sn samples irradiated under gamma-ray source and 10FFF x-ray beam in this study, and Zn film samples under 6MeV electron beam from [\cite{niraula2016}].  \label{fig:DoseRateDep}}.
\end{figure}

The above paragraph interpretation predicts a maximum number of whiskers on the fully exposed sample, a minimum on the non-exposed (control) one, and then intermediate comparable numbers on the half exposed and screened samples.  These predictions are mostly confirmed by the data in Fig. \ref{fig:stats10FFF} with the exception that the exposed and shielded halves are not on par. There are several conceivable explanations for the observed differences between the shielded and unshielded halves. For example, irradiation can generate additional free carriers unaccounted for in our theoretical analysis of Sec. \ref{sec:theory}, or some effects of structural damages under high energy photons can contribute to whiskering.

The significance of our results is twofold. For one, they experimentally verify the electrostatic model of tin whisker development thus pointing at possible `electrostatic ways' of their mitigation, such as developing electrically active surface treatments, enforcing system grounding, etc. On the other hand, our results bring up possible issues with tin whisker developments in radiation-active environment, such as space applications or electron accelerators, including medical linacs. Finally they point at a possible accelerated testing for whisker propensity using ionizing radiation.

For the latter venue, we quantify the effect of radiation induced electric field on whisker growth using the whisker creation rate defined as
\begin{equation}\label{eq:R}
R=\frac{{\rm number\ of \  whiskers}}{{\rm area}\times {\rm time}}=\frac{{\rm whisker \ density}}{{\rm time}}\end{equation}	
Here, we distinguish between $R_{\rm spon}$ - spontaneous creation rate, with no external field applied, and $R_{\rm stim}$ - stimulated by applied external electric field, generated under gamma- or x-ray irradiation. We define the acceleration factor
	\begin{equation}\label{eq:a}a\equiv R_{\rm stim}/R_{\rm spon},\end{equation}	
which can be numerically estimated as $a\sim 55$ for the central strips receiving 30 kGy under gamma-ray irradiation. This value is very close to $a \sim 52$ obtained in our previous experiments under the same source. \cite{Killefer} For the x-ray irradiation the acceleration factor is several times higher, getting as high as $\sim 246$ for fully irradiated sample. Comparing the latter result with those previously measured for tin and zinc samples under 6 MeV electron beams, \cite{vasko2015,niraula2016} we point out similar values of acceleration factor observed under x-ray beam irradiation here. Summarizing all mentioned studies we note an evident dose rate dependence illustrated in Fig. \ref{fig:DoseRateDep} for the three ionizing radiation sources (disregarding the differences in their beam quality). We attribute the dose rate dependence to the radiation charging capacity, resulting in lower sustained electric fields during irradiation with lower dose rate sources. The final achieved total dose levels are close to 30kGy in all three experiments ($\sim 26$ kGy under the 20-hour electron beam irradiation of reference \cite{vasko2015,niraula2016}).

From a practical standpoint that acceleration means growing whiskers under high-energy photon (gamma- or x-ray) or electron sources should take one-two weeks rather than years.

\section{Conclusions}\label{sec:concl}
Based on results of our experiments and related theoretical considerations, the following conclusions about whisker physics and its practical implications can be made:\\
(1) We experimentally observed the accelerated Sn
whisker growth under non-destructive gamma-ray and x-ray irradiation
and determined the characteristic range of radiation doses 20-30 kGy, for which that effect becomes significant.\\
(2) We were able to change the radiation induced whisker growth by electrically disconnecting the parts of our experimental setup thus demonstrating the electrostatic nature of the accelerated whisker development.\\
(3) By observing the delayed kinetics of radiation induced whisker development, we conclude that it affects mostly the whisker nucleation stage.\\
(4) The observed acceleration factors make the ionizing radiation a potential non-destructive and readily implementable
accelerated life testing tool. The observed dose rate dependance further confirms the electrostatic nature of the whisker growth acceleration.

\appendix
\section{$\delta$-function}\label{append}

One property of the $\delta$-function is obvious:
\begin{equation}
\lim _{H\rightarrow 0}\frac{1}{2\pi H}\ln\frac{z^2+H^2}{z^2}=0 \quad {\rm when}\quad  z\neq 0.\end{equation}
Integrating by parts renders its other property:
\begin{eqnarray}\label{eq:unity}
&&\lim _{H\rightarrow 0}\frac{1}{2\pi H}\left[z\ln\frac{z^2+H^2}{z^2}|_{-\infty}^{\infty}+2\int _{-\infty}^{\infty}\frac{H^2}{H^2+z^2}dz\right]=\nonumber \\
&&\lim _{H\rightarrow 0}\frac{1}{2\pi H}2H^2\frac{1}{H}\arctan\frac{z}{H}| _{-\infty}^{\infty}=1.\end{eqnarray}
\section*{Acknowledgement}
Microscopy for this study was done with equipment at the
Center for Materials and Sensor Characterization center at the
University of Toledo.

\end{document}